\documentclass[conference]{IEEEtran}

\IEEEoverridecommandlockouts

\usepackage{cite}
\usepackage{overpic}
\usepackage{amsmath,amssymb,amsfonts}
\usepackage{graphicx}
\usepackage{textcomp}
\usepackage{xcolor}
\usepackage{float}
\usepackage{amsthm}
\usepackage{graphicx}
\usepackage{epstopdf}
\usepackage{subfigure}
\usepackage{adjustbox}
\usepackage{amsmath,bm}
\usepackage{amsfonts}
\usepackage{amssymb}
\usepackage{color}
\usepackage{multirow}
\usepackage{multicol}
\usepackage{soul,xcolor}
\usepackage{algorithm}
\usepackage{algpseudocode}%

\theoremstyle{plain}

\newcommand{\vect}[1]{\mathbf{#1}}

\def\Ttran{\mbox{\tiny $\mathrm{T}$}}
\def\imagunit{\mathsf{j}} 

\newcommand{\argmin}[1]{{\underset{{#1}}{\mathrm{arg\,min}}}}

\begin{document}

\title{Joint Impact of ADC and Fronthaul Quantization in Cell-Free Massive MIMO-OFDM Uplink 

}

\author{\IEEEauthorblockN{ Özlem Tuğfe Demir\IEEEauthorrefmark{1} and Emil Björnson\IEEEauthorrefmark{2}\thanks{This work was carried out within the scope of the project 122C149 – Intelligent End-to-End Design of Energy-Efficient and Hardware Impairments-Aware Cell-Free Massive MIMO for Beyond 5G. \"O. T. Demir was supported by the 2232-B International Fellowship for Early Stage Researchers Programme funded by the Scientific and Technological Research Council of Türkiye (TÜBİTAK). E.~Bj\"ornson was supported by the Knut and Alice Wallenberg Foundation through a WAF grant.}}
\IEEEauthorblockA{ {$^*$Department of Electrical and Electronics Engineering, Bilkent University, Ankara, Turkiye
		}\\{$^\dagger$Department of Communication Systems, KTH Royal Institute of Technology, Stockholm, Sweden
		} 
   \\
		\IEEEauthorblockA{E-mail: \IEEEauthorrefmark{1}ozlemtugfedemir@bilkent.edu.tr, \IEEEauthorrefmark{2}emilbjo@kth.se}
}
}

\maketitle

\begin{abstract}
In the uplink of a cell-free massive MIMO system, quantization affects performance in two key domains: the time-domain distortion introduced by finite-resolution analog-to-digital converters (ADCs) at the access points (APs), and the fronthaul quantization of signals sent to the central processing unit (CPU). Although quantizing twice may seem redundant, the ADC quantization in orthogonal frequency-division duplex (OFDM) systems appears in the time domain, and one must then convert to the frequency domain, where quantization can be applied only to the signals at active subcarriers. This reduces fronthaul load and avoids unnecessary distortion, since the ADC output spans all OFDM samples while only a subset of subcarriers carries useful information.

While both quantization effects have been extensively studied in narrowband systems, their joint impact in practical wideband OFDM-based cell-free massive MIMO remains largely unexplored. This paper addresses the gap by modeling the joint distortion and proposing a fronthaul strategy in which each AP processes the received signal to reduce quantization artifacts before transmission. We develop an efficient estimation algorithm that reconstructs the unquantized time-domain signal prior to fronthaul transmission and evaluate its effectiveness. The proposed design offers new insights for implementing efficient, quantization-aware uplink transmission in wideband cell-free architectures.
\end{abstract}

\section{Introduction}

Cell-free massive MIMO is a next-generation wireless architecture in which a large number of distributed access points (APs) collaboratively serve multiple user equipments (UEs) over the same time-frequency resources, without relying on traditional cell boundaries \cite{Ngo2017b,cell-free-book,ngo2024ultradense}. This dense and cooperative deployment enables seamless connectivity, boosting both spectral and energy efficiency compared to cellular technology.

To ensure cost-efficiency and scalability, each AP is typically equipped with a small number of antennas and low-cost radio-frequency (RF) components. However, such economical implementations inevitably introduce hardware impairments \cite{8891922}. One common approach to reducing RF chain costs is the use of low-resolution analog-to-digital converters (ADCs), which, while cost-effective, introduce significant quantization distortion that degrades signal quality. Furthermore, the fronthaul link—responsible for transmitting the received data from the APs to the central processing unit (CPU)—can also become a source of distortion when the data is quantized with limited resolution.

While prior studies have considered either ADC impairments or fronthaul quantization independently \cite{8756265,8678745} or jointly \cite{8891922,9123382}, they have focused on narrowband models and fail to fully capture how the quantization affects the time-domain signals in wideband systems based on orthogonal frequency division multiplexing (OFDM). This work addresses this gap by analyzing the uplink of a wideband cell-free massive MIMO system employing OFDM, with a focus on the joint impact of ADC distortion and fronthaul quantization.  

Unlike in narrowband systems, quantization in wideband OFDM-based architectures occurs in different domains: ADC quantization is performed in the time domain, while fronthaul quantization is applied in the frequency domain. The latter approach reduces fronthaul data rate requirements, as only the subcarriers carrying data or pilot symbols are quantized, as opposed to the full set of time-domain samples. Although the number of used subcarriers can still represent a significant portion of the DFT size (often 50–75\%), additional overhead in the time domain—due to cyclic prefix and potential oversampling for digital filtering—makes time-domain quantization less efficient. Transforming the signal to the frequency domain eliminates this redundancy. A feasible yet suboptimal method is to quantize the signal both before (due to ADCs) and after the DFT (due to limited fronthaul resolution), resulting in a double quantization effect that degrades signal fidelity.

To overcome this limitation, we propose a novel signal processing strategy that reconstructs the clean time-domain signal prior to fronthaul transmission, thereby mitigating the cumulative effects of double quantization. By exploiting the structure of the symbol constellation, we formulate the problem in a manner that enables an efficient iterative algorithm based on alternating direction method of multipliers (ADMM) to estimate the unquantized signal using the ADC outputs and known pilot information. The proposed method offers new insights into quantization-aware signal processing.

\section{System Model and Problem Formulation}

The uplink of a cell-free massive MIMO-OFDM system is considered. There are $L$ APs and $K$ user equipments (UEs), distributed arbitrarily within the network area. All UEs have a single antenna while each AP is equipped with $N$ antennas. The complex baseband equivalent frequency-selective channel from each UE to each AP is modeled as a finite impulse response (FIR) filter with $R$ equally sample-spaced taps.  The sampling period is $T_s=1/B$ where $B$ is the total bandwidth of $M$ OFDM subcarriers with $B/M$ subcarrier spacing. The set of all subcarriers is denoted by $\mathcal{M}=\{0,\ldots,M-1\}$.

The $r$-th tap of the channel from UE $k$ to AP $l$ is denoted by ${\bf h}_{kl}[r]\in \mathbb{C}^{N}$, for $r=0,\ldots,R-1$. The number of subcarriers satisfies $M>R$, and there are $M+R-1$ samples in each OFDM symbol with a cyclic prefix (CP) length of $R-1$. 

Let us focus on one of the OFDM symbols that is allocated for uplink data transmission. We denote the set of subcarriers used for data transmission by $\mathcal{M}_{\rm used}$, where $\overline{M}=|\mathcal{M}_{\rm used}|$ is usually less than the total number of subcarriers $M$, which is also the DFT size. The other subcarriers in the set $\mathcal{M}\setminus \mathcal{M}_{\rm used}$ are either used for pilot transmission or kept null. The frequency-domain uplink symbol of UE $k$ at the $m$th subcarrier is given by $\overline{s}_k[m]$. We will use the same notation $\overline{s}_k[m]$ to denote the pilot subcarriers of UE $k$ or zero signal at the null subcarriers.  The baseband equivalent signal transmitted from UE $k$ at the $q$th time-domain sample is given by the $M$-point inverse DFT of the sequence $\overline{s}_k[m]$, i.e.,
\vspace{-2mm}
\begin{align}
s_k[q] =
\frac{1}{\sqrt{M}}\sum_{m=0}^{M-1}\overline{s}_k[m]e^{\imagunit \frac{2\pi q m}{M}}, 
\end{align}
where $q=-(R-1),\ldots,M-1$, and we assume that a CP of length $R-1$ is appended to the time-domain samples. The baseband equivalent of the distortion-free received signal at the $N$ antennas of AP $l$ is given as
\vspace{-2mm}
\begin{align}
\vect{y}_{l}'[q]=\underbrace{\sum_{k=1}^K\sum_{r=0}^{R-1}\vect{h}_{kl}[r]\sqrt{p_k}s_k[q-r]}_{\triangleq \vect{z}_l[q]} + \vect{w}_l[q],  \label{eq:received-data}
\end{align}
for $q=0,\ldots,M-1$. Let $Q_{\rm ADC}(\cdot)$ denote the quantization function of the ADCs with $b^{\rm ADC}$ bits and $D^{\rm ADC}=2^{b^{\rm ADC}}$ quantization levels, $l_1,\ldots,l_{D^{\rm ADC}}$,
and the quantization function is given by
\begin{align} \label{eq:quantization-function}
Q_{\rm ADC}(x)=l_d \quad \text{if } x\in[ \upsilon_{d-1}, \upsilon_d ), \quad d=1,\ldots,D^{\rm ADC},
\end{align}
where $-\infty=\upsilon_0<\upsilon_1<\cdots<\upsilon_{D^{\rm ADC}-1}<\upsilon_{D^{\rm ADC}}=\infty$ are the thresholds. By applying the scalar quantization $Q_{\rm ADC}(\cdot)$ independently to the real and imaginary parts of  $\vect{y}_l'[q]$, we obtain the complex-valued quantized signal $\vect{y}_l[q] \in \mathbb{C}^{N}$ as
\begin{align} \label{eq:tilder}
\vect{y}_l[q]=Q_{\rm ADC}\left(\Re(\vect{y}_l'[q])\right)+\imagunit Q_{\rm ADC}\left(\Im(\vect{y}_l'[q])\right).
\end{align}

The aim of the fronthaul transmission is to determine the frequency-domain quantized signals $\overline{\vect{z}}_l[m]$, for $m\in \mathcal{M}_{\rm used}$ to be transmitted from AP $l$ to the CPU so that the mean square distance to the DFT of the clean and distortion-free signal $\vect{z}_l[q]$ is reduced as much as possible. To achieve this, we propose an efficient algorithm that estimates 
$\vect{z}_l[q]$ by leveraging both the ADC quantization levels and the modulation structure of the UE symbols. Subsequently, the classical Lloyd–Max algorithm can be applied to the estimated unquantized signal to compress it prior to fronthaul transmission. In the next section, we will develop an ADMM-based algorithm to estimate the clean unquantized signal $\vect{z}_l[q]$ using the quantized signals $\vect{y}_l[q]$ for a particular AP $l$.

\section{Unquantized Signal Estimation with ADMM}

We first write the clean, unquantized signal as
\begin{align}
\vect{z}_{l}[q]=\sum_{k=1}^K\sum_{r=0}^{R-1}\vect{h}_{kl}[r]\sqrt{p_k}\frac{1}{\sqrt{M}}\sum_{m=0}^{M-1}\overline{s}_k[m]e^{\imagunit \frac{2\pi (q-r) m}{M}}.  \label{eq:received-data}
\end{align}
We assume that all subcarriers in
$\mathcal{M}\setminus\mathcal{M}_{\rm used}$ either carry deterministic pilot
symbols known at AP $l$ or are intentionally left empty (zero symbols). Under the assumption of perfect channel state information (CSI), the part of the above signal corresponding to pilot or null subcarriers is already known at AP $l$. We will denote this signal by $\vect{p}_l[q]$ and it is given as
\begin{align}
\vect{p}_{l}[q]=\sum_{k=1}^K\sum_{r=0}^{R-1}\vect{h}_{kl}[r]\frac{\sqrt{p_k}}{\sqrt{M}}\sum_{m\in \mathcal{M}\setminus \mathcal{M}_{\rm used}}\overline{s}_k[m]e^{\imagunit \frac{2\pi (q-r) m}{M}}.  \label{eq:received-data2}
\end{align}
Now, to construct a matrix-vector multiplication relation between the data symbols and the unquantized signals, we define 
\begin{align}
&\overline{\mathbf{S}}
=
\begin{bmatrix}
\overline{\mathbf{s}}_{1} &
\overline{\mathbf{s}}_{2} &
\cdots &
\overline{\mathbf{s}}_{K}
\end{bmatrix}
\in \mathbb{C}^{\overline{M}\times K}, 
\end{align}
where $\overline{\vect{s}}_k$ is given as
\begin{align}
    \overline{\vect{s}}_k = \left[ \overline{s}_k[m_1], \ \ldots, \ \overline{s}_k[m_{\overline{M}}] \right]^{\Ttran} \in \mathbb{C}^{\overline{M}}.
\end{align}
Here, $m_1,\ldots,m_{\overline{M}}$ are the indices corresponding to used subcarriers. Moreover, we define
\begin{align}
&\mathbf{d}_{q-r}
=
\frac{1}{\sqrt{M}}
\begin{bmatrix}
e^{\jmath \frac{2\pi(q-r)m_1}{M}} &
\cdots &
e^{\jmath \frac{2\pi(q-r)m_{\overline{M}}}{M}}
\end{bmatrix}^{\Ttran}
\in \mathbb{C}^{\overline{M}}, \\
& \mathbf{H}_{l}[r]
=
\begin{bmatrix}
\mathbf{h}_{1l}[r] &
\mathbf{h}_{2l}[r] &
\cdots &
\mathbf{h}_{Kl}[r]
\end{bmatrix}
\in \mathbb{C}^{N\times K}, \\
&\mathbf{P}^{1/2}
=
\operatorname{diag}\!\left(\sqrt{p_1},\sqrt{p_2},\ldots,\sqrt{p_K}\right), \\
& \mathbf{A}_{l}[q]
=
\sum_{r=0}^{R-1}
\left(
\mathbf{H}_{l}[r]\,
\mathbf{P}^{1/2}
\otimes
\mathbf{d}_{q-r}^{\Ttran}
\right)
\in \mathbb{C}^{N\times (K\overline{M})}.
\end{align}
Then, we can write \eqref{eq:received-data} as
\begin{align}
& \mathbf{z}_{l}[q]
=
\mathbf{A}_{l}[q]\,
\mathrm{vec}\!\left(\overline{\mathbf{S}}\right)+\vect{p}_l[q].
\end{align}
Concatenating all the time-domain samples, we obtain
\begin{align}
    \vect{z}_l 
    &= 
    \begin{bmatrix}
        \vect{z}_l[0] \\
        \vect{z}_l[1] \\
        \vdots \\
        \vect{z}_l[M-1]
    \end{bmatrix}
    \in \mathbb{C}^{NM}, \quad \vect{p}_l 
    &= 
    \begin{bmatrix}
        \vect{p}_l[0] \\
        \vect{p}_l[1] \\
        \vdots \\
        \vect{p}_l[M-1]
    \end{bmatrix}
    \in \mathbb{C}^{NM}.
    \end{align}
    Then, by defining
    \begin{align}
    \vect{x} 
    &= \mathrm{vec}\left(\overline{\vect{S}}\right) 
    \in \mathbb{C}^{K\overline{M}}, \quad  \vect{A}_l 
    =
    \begin{bmatrix}
        \vect{A}_l[0] \\
        \vect{A}_l[1] \\
        \vdots \\
        \vect{A}_l[M-1]
    \end{bmatrix}\mathbb{C}^{NM \times K\overline{M}},
    \end{align}
    we obtain the linear relation
    \begin{align}
    \vect{z}_l 
    &= \vect{A}_l \vect{x}+\vect{p}_l. \label{eq:linear}
\end{align}

Our next goal is to express the problem of estimating the unquantized signal in real-valued form. To this end, we define the real vector using $\vect{x} 
    = \mathrm{vec}\left(\overline{\vect{S}}\right)$
\begin{align}
    \vect{x_r}
    =
    \begin{bmatrix}
        \Re\left(\vect{x}\right) \\
        \Im\left(\vect{x}\right)
    \end{bmatrix}
    \in \mathbb{R}^{2K\overline{M}}.
\end{align}
Further defining
\begin{align}
&\vect{z_r}_l = \begin{bmatrix} \Re\left(\vect{z}_l\right) \\ \Im\left(\vect{z}_l\right) \end{bmatrix}, \quad \vect{p_r}_l = \begin{bmatrix} \Re\left(\vect{p}_l\right) \\ \Im\left(\vect{p}_l\right) \end{bmatrix}, \\
&
\vect{A_r}_l = \begin{bmatrix} \Re\left(\vect{A}_l\right) &  -\Im\left(\vect{A}_l\right) \\
\Im\left(\vect{A}_l\right) & \Re\left(\vect{A}_l\right)
\end{bmatrix},
\end{align}
we can write \eqref{eq:linear} in a real-valued form as
\begin{align}
    \vect{z_r}_l = \vect{A_r}_l\vect{x_r}+\vect{p_r}_l.
\end{align}
Our method is to estimate the unquantized signal $\vect{z_r}_l$ by detecting the unknown symbols $\vect{x_r}$, where $\vect{A_r}_l$ is known at AP $l$ from CSI.

By extending the approach in \cite{demirWiopt} to incorporate the ADC codebook and the presence of the pilot signal $\vect{p_r}_l$, the symbol detection task can be formulated as an optimization problem. We treat $\vect{z_r}_l$ and $\vect{x_r}$ as optimization variables, and our goal is to minimize the Euclidean distance between $\vect{z_r}_l$ and the quantized signal 
\begin{align}
  \vect{y_r}_l = \begin{bmatrix} \Re\left(\vect{y}_l\right) \\
 \Im\left(\vect{y}_l\right) \end{bmatrix} \in \mathbb{R}^{2NM},
 \end{align}
 where
 \begin{align}
     \vect{y}_l=\begin{bmatrix} \vect{y}_l[0] \\ \vdots \\ \vect{y}_l[M-1]\end{bmatrix}\in \mathbb{C}^{NM}.
 \end{align}

Given the quantized signal $\vect{y}_l[q]$ in \eqref{eq:tilder}, each pre-ADC sample $\left[\vect{y_r}_l'\right]_n$, which is the $n$-th element of $\vect{y_r}_l'$ (defined in a manner analogous to the preceding expression),  must lie in the
interval corresponding to the observed quantization level. We express
this as
\begin{align}
    \nu_{\mathrm{low},n}
    \;\leq\;
    \left[\vect{y_r}_l'\right]_n
    \;<\;
    \nu_{\mathrm{upp},n},
    \qquad n=1,\ldots,2NM,
    \label{eq:q-interval}
\end{align}
where $\nu_{\mathrm{low},n}=\upsilon_{d-1}$ and
$\nu_{\mathrm{upp},n}=\upsilon_{d}$ if the corresponding element of the quantized vector is $l_d$. In the optimization problem we propose, we will replace the inequalities in \eqref{eq:q-interval} by 
\begin{align}
    \nu_{\mathrm{low},n}
    \;\leq\;
    \left[\vect{z_r}_l\right]_n
    \;<\;
    \nu_{\mathrm{upp},n},
    \qquad n=1,\ldots,2NM,
    \label{eq:q-interval2}
\end{align}
where we neglect the impact of noise and assume that the clean signal lies within the same intervals as the noisy pre-ADC signal.
\subsection{ADMM formulation}

We now formulate an ADMM-friendly optimization problem by
introducing a copy of the unquantized signal and the symbol vector. Let
\begin{align}
    \vect{t}_l = \vect{z_r}_l,
    \qquad
    \vect{b} = \vect{x_r}.
\end{align}
The first relation will enable enforcing the projection onto the quantization intervals, while the second will be useful in imposing the discrete symbol constellation constraints. 

We assume the data signals belong to a finite constellation set. Let $\mathcal{O}^{\Re}$ and $\mathcal{O}^{\Im}$ denote the sets of all possible real and imaginary parts of a constellation point. For QAM constellations, the real and imaginary parts of each symbol are independent. On the other hand, for PSK constellations except QPSK, there is a dependency between the real and imaginary parts of each symbol. However, we will neglect this dependency for simplicity, and we obtain the following non-convex problem with the constellation  constraints:
\begin{subequations}\label{eq:problem}
    \begin{align}
\underset{\vect{z_r}_l,\;\vect{t}_l,\;\vect{x_r},\;\vect{b}}{\text{minimize}}
    \quad
        & \left\Vert \vect{z_r}_l-\vect{y_r}_l\right\Vert^2
        \label{prob:main}\\
    \text{subject to}\quad
        & \vect{t}_l = \vect{z_r}_l, \label{cons:t=z}\\
        & \vect{b} = \vect{x_r}, \label{cons:b=Az}\\
        & \vect{z_r}_l=\vect{A_r}_l\vect{x_r}+\vect{p_r}_l, \label{eq:linear-equality} \\
        & \nu_{\mathrm{low},n}
    \;\leq\;
   t_{ln}
    \;<\;
    \nu_{\mathrm{upp},n},
    \quad n=1,\ldots,2NM,
            \label{cons:interval}\\
        & b_{i}\in \mathcal{O}^{\Re}, \ \ \ b_{K\overline{M}+i}\in \mathcal{O}^{\Im}, \ \ \  i=1,\ldots,K\overline{M},
            \label{cons:binary}
\end{align}
\end{subequations}
where $t_{ln}$ is the $n$-th entry of $\vect{t}_l$ and $b_i$ is the $i$-th entry of $\vect{b}$. The constraints \eqref{cons:interval} capture the general $b^{\rm ADC}$-bit ADC
quantization structure, while \eqref{cons:binary} enforces the modulation
constraints. This reformulation, which introduces auxiliary copies of the optimization variables, enables the derivation of closed-form ADMM updates.

\subsection{ADMM updates}

We apply ADMM in scaled form to \eqref{eq:problem}. Let
$\vect{u}_1\in\mathbb{R}^{2NM}$,
$\vect{u}_2\in\mathbb{R}^{2K\overline{M}}$, and $\vect{u}_3\in\mathbb{R}^{2NM} $ denote the scaled dual variables
associated with the equality constraints \eqref{cons:t=z},
\eqref{cons:b=Az}, and $\eqref{eq:linear-equality}$, respectively. At iteration $r$, the update equations are
\begin{align}
   \vect{z_r}_l^{(r+1)}
    &\leftarrow
    \underset{\vect{z_r}_l}{\arg\min}
        \;
        \left\Vert \vect{z_r}_l-\vect{y_r}_l\right\Vert^2
        + \rho
        \bigl\|
            \vect{t}_l^{(r)} - \vect{z_r}_l + \vect{u}_1^{(r)}
        \bigr\|_2^2
        \nonumber\\
        &\hspace{8mm}
        + \rho
        \bigl\|
            \vect{z_r}_l - \vect{A_r}_l\vect{x_r}^{(r)} -\vect{p_r}_l+ \vect{u}_3^{(r)}
        \bigr\|_2^2
        \nonumber\\[1mm]
    &\hspace{-12mm}=
    \frac{1}{1+2\rho}\Bigl( \vect{y_r}_l+\rho
        \left(\vect{t}_l^{(r)}+\vect{u}_1^{(r)}+\vect{A_r}_l\vect{x}_r^{(r)}+\vect{p_r}_l-\vect{u}_3^{(r)}\right)
        \Bigr),
    \label{ADMM:z-update}
\end{align}
where $\rho$ is the penalty parameter. The update equation for $\vect{b}$ is given as
\begin{align}
\vect{b}^{(r+1)} &\leftarrow
    \underset{\vect{b}}{\arg\min}
        \;  \rho\left\Vert \vect{b}-\vect{x_r}^{(r)}+\vect{u}_2^{(r)} \right\Vert^2 \nonumber\\
   &   \hspace{-6mm}  \text{subject to} \; b_{i}\in \mathcal{O}^{\Re}, \ \ \ b_{K\overline{M}+i}\in \mathcal{O}^{\Im}, \ \ \  i=1,\ldots,K\overline{M}.
\end{align}
A closed-form solution can be found by projecting the entries of $\vect{x_r}^{(r)}-\vect{u}_2^{(r)}$ onto the corresponding modulation constraint set and given as
\begin{align} &b_{i}^{(r+1)}=\Pi_{s}^{\Re}\Big(\left[\vect{x_r}^{(r)}-\vect{u}_2^{(r)}\right]_i\Big), \\
& b_{K\overline{M}+i}^{(r+1)}=\Pi_{s}^{\Im}\Big(\left[\vect{x_r}^{(r)}-\vect{u}_2^{(r)}\right]_{K\overline{M}+i}\Big),
\end{align}
for $i=1,\ldots,K\overline{M}$, where the projection functions are given as
\begin{align}
    \Pi_{s}^{\Re} \big(x\big)= \argmin{s\in \mathcal{O}^{\Re}} \left\vert x-s\right\vert, \quad   \Pi_{s}^{\Im} \big(x\big)= \argmin{s\in \mathcal{O}^{\Im}} \left\vert x-s\right\vert.
\end{align}
Proceeding with the remaining updates, we obtain
\begin{align}
    \vect{t}_l^{(r+1)}
    &\leftarrow
     \underset{\vect{t}_l}{\arg\min}
        \;
        \rho
        \bigl\|
            \vect{t}_l - \vect{z_r}_l^{(r+1)} + \vect{u}_1^{(r)}
        \bigr\|_2^2
        \nonumber\\
    &\hspace{-8mm}\text{subject to}\;
        \nu_{\mathrm{low},n}
    \;\leq\;
    t_{ln}
    \;<\;
    \nu_{\mathrm{upp},n},
    \qquad n=1,\ldots,2NM.
\end{align}
This yields the element-wise projection
\begin{align}
    t_{ln}^{(r+1)}
    =
    \Pi_{\bigl[\nu_{\mathrm{low},n},\,\nu_{\mathrm{upp},n}\bigr]}
    \!\Bigl(
        \left[\vect{z_r}_l^{(r+1)} - \vect{u}_1^{(r)}\right]_n
    \Bigr),
    \label{ADMM:t-update}
\end{align}
where
\begin{align}
    \Pi_{[a,b]}(x)
    =
    \begin{cases}
        a, & x < a,\\
        x, & a \leq x \leq b,\\
        b, & x > b.
    \end{cases}
\end{align}
\begin{align}
\vect{x_r}^{(r+1)} &\leftarrow
    \underset{\vect{x_r}}{\arg\min}
        \;  \rho\left\Vert \vect{b}^{(r+1)}-\vect{x_r}+\vect{u}_2^{(r)} \right\Vert^2 \\&\hspace{8mm}
        + \rho
        \bigl\|
            \vect{z_r}_l^{(r+1)} - \vect{A_r}_l\vect{x_r} -\vect{p_r}_l+ \vect{u}_3^{(r)}
        \bigr\|_2^2 \nonumber \\
  &  = \left(\vect{I}_{2K\overline{M}}+\vect{A_r}_l^{\Ttran}\vect{A_r}_l\right)^{-1} \nonumber\\
  &\hspace{-8mm}\cdot\left( \vect{b}^{(r+1)}+\vect{u}_2^{(r)}+\vect{A_r}_l^{\Ttran}\left(\vect{z_r}_l^{(r+1)}-\vect{p_r}_l+\vect{u}_3^{(r)}\right)\right).
\end{align}
Finally, we obtain the dual variable updates
\begin{align}
    \vect{u}_1^{(r+1)}
    &= \vect{u}_1^{(r)}+\vect{t}_l^{(r+1)} - \vect{z_r}_l^{(r+1)},
       \label{ADMM:u1-update}\\
    \vect{u}_2^{(r+1)}
    &= \vect{u}_2^{(r)}
       + \vect{b}^{(r+1)} - \vect{x_r}^{(r+1)}, 
       \label{ADMM:u2-update} \\
\vect{u}_3^{(r+1)}
    &= \vect{u}_3^{(r)} +\vect{z_r}_l^{(r+1)}-\vect{A_r}_l\vect{x_r}^{(r+1)}-\vect{p_r}_l.
\end{align}

\section{Numerical Experiments}
In this section, we quantify the quantization error of the frequency-domain
uplink signals transmitted through the fronthaul channel in a cell-free massive
MIMO system with low-resolution ADCs. We consider a single representative AP
randomly deployed in a $500 \times 500$\,m$^2$ area with uniform distribution. The AP is equipped with
$N=4$ antennas, and $K=8$ UEs are randomly positioned within the same area.
The path loss in decibel scale follows the Urban Microcell Street Canyon model for a carrier frequency of $f_c = 7.5$\,GHz and
and is given by
$-49.9 - 31.9\log_{10}(d) + \mathcal{N}(0,8.2^2)$,
where $d$ is the distance between two nodes in meters and $8.2$\,dB is the standard deviation of the lognormal
shadow fading \cite[Table~7.4.1-1]{3GPP5G}. A height difference of $10$\,m is
assumed between the AP and the UEs. The total number of OFDM subcarriers is
$M=256$, with $R=6$ channel taps and a subcarrier spacing of $60$\,kHz. The
noise variance is computed for a bandwidth of $B = 256 \cdot 60\,\text{kHz} =
15.36$\,MHz and a noise figure of $7$\,dB, yielding $\sigma^2 = -95.14$\,dBm.
The Saleh--Valenzuela model \cite{Saleh1987} is used to generate the power delay
profile with $\Gamma = \gamma = 2$ and five clusters
\cite[Eq.~(7.52)]{bjornson2024introduction}. The $N=4$ AP antennas are arranged
in a $2 \times 2$ uniform planar array. The uplink transmit power is fixed to
$p_k = 0.1$\,W for all UEs. The Lloyd--Max algorithm with symmetry around zero
is used to design the ADC quantization levels. The ADMM algorithm is executed
with $20$ iterations and parameter $\rho = 10$.

Out of the total $M=256$ subcarriers, $\overline{M}=64$ subcarriers are
allocated to data transmission. These data-carrying subcarriers are chosen to be
\emph{uniformly spaced} across the index range $[0,M-1]$, such that the spacing
between consecutive data subcarriers is constant. Among the remaining subcarriers, $64$
randomly selected ones carry unit-power pilots (it is simply selected as $1$), and the remaining $128$
subcarriers are nulled (i.e., no transmission takes place).

We evaluate two schemes:  
(i)~the proposed approach, where the ADMM algorithm is run at the AP to estimate
the clean time-domain signal before applying the Lloyd--Max quantizer to the
active subcarriers, and  
(ii)~a benchmark scheme that directly quantizes the ADC-distorted frequency-domain
signals using the Lloyd--Max algorithm. Both schemes employ symmetric
quantization levels.

We consider three different fronthaul quantization resolutions per real
dimension: $b_{\rm frt}=2$, $b_{\rm frt}=4$, and $b_{\rm frt}=\infty$, where
$b_{\rm frt}=\infty$ corresponds to unquantized fronthaul transmission. The
horizontal axis of the figures varies the ADC resolution $b_{\rm ADC}$ per real
dimension.

\begin{figure}[t] 
    \centering
        \includegraphics[width=0.5\textwidth, trim=0.2cm 0.2cm 1cm 0.2cm, clip]{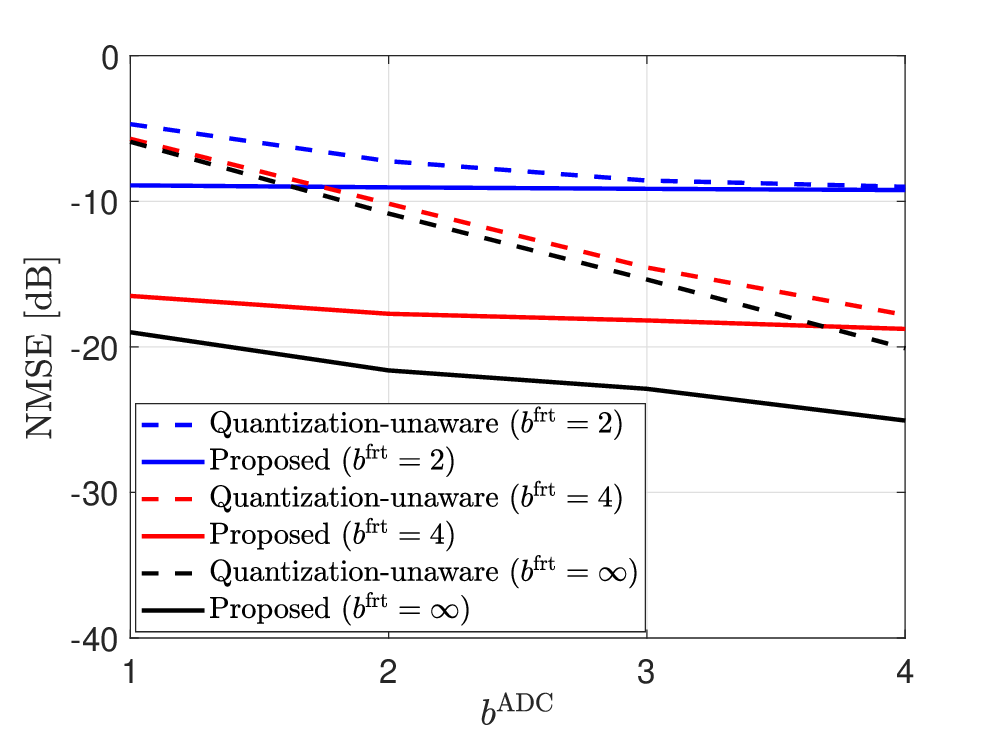} 
        \caption{NMSE of the reconstructed frequency-domain uplink signal for QPSK modulation as a function of the ADC resolution.}
        \label{fig:1}
 \end{figure}

Fig.~\ref{fig:1} shows the normalized mean-square error (NMSE) between the
clean, noiseless, unquantized frequency-domain signal at the active subcarriers
and its quantized version for QPSK modulation. The results are averaged over
random AP and UE locations, giving higher weight to realizations with larger
received signal power---a meaningful feature since such APs contribute more
strongly to the overall data detection at the CPU. The proposed method
consistently outperforms the quantization-unaware benchmark. By
\emph{quantization-unaware}, we refer to the fact that the benchmark does not
account for the ADC quantization structure, whereas the proposed method exploits
both the modulation constraints and the ADC quantization partitions when
reconstructing the clean signal.

As the ADC resolution increases, the performance gap between the methods
narrows, particularly when the fronthaul resolution is limited. However, a
substantial improvement persists even for $b_{\rm frt} = \infty$ at
$b_{\rm ADC} = 4$ bits. Notably, the proposed algorithm can effectively recover
the signal even with very low ADC resolutions.

\begin{figure}[t] 
    \centering
        \includegraphics[width=0.5\textwidth, trim=0.2cm 0.2cm 1cm 0.2cm, clip]{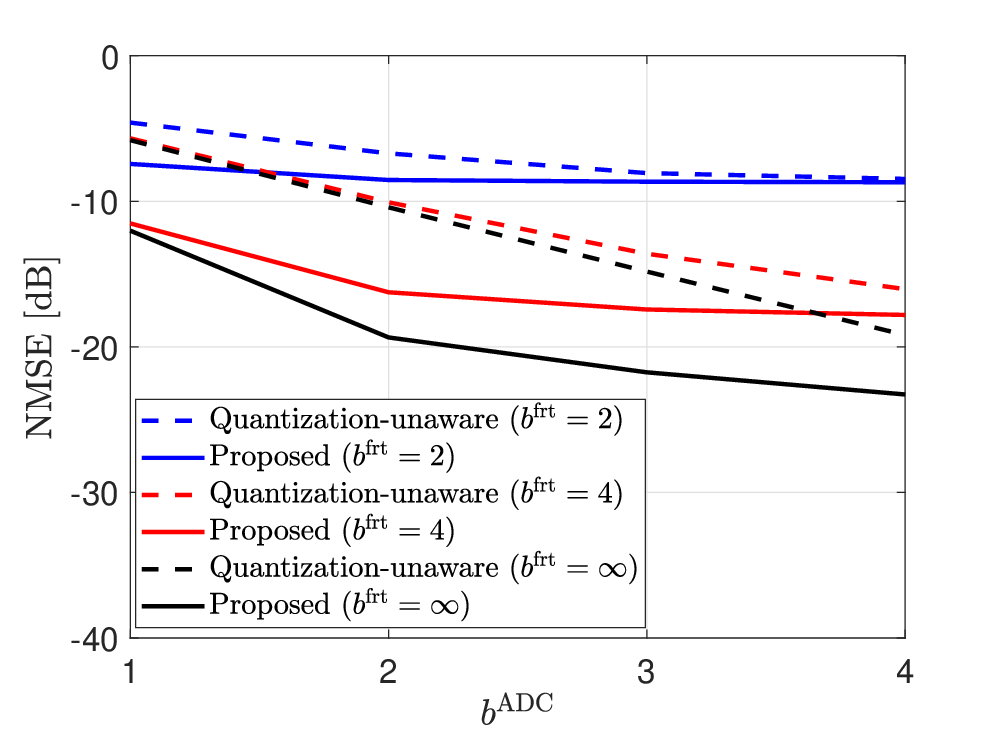} 
        \caption{NMSE of the reconstructed frequency-domain uplink signal for 16-QAM modulation as a function of the ADC resolution.}
        \label{fig:2}
 \end{figure}

In Fig.~\ref{fig:2}, we repeat the experiment for 16-QAM. As expected, the NMSE
values are generally higher due to the increased constellation size, and the
performance degradation is more visible at low ADC resolutions. Nonetheless, the
proposed scheme still provides a clear improvement over the benchmark.

Although the AP is equipped with only $4$ antennas, the signals of $8$ UEs can
be detected with sufficient accuracy to yield high-quality estimates of the
clean received signal. We emphasize that these local estimates are not the final
decoded symbols; rather, the CPU performs the ultimate data detection after
collecting the quantized fronthaul signals. Developing an efficient decoding
strategy at the CPU based on these quantized observations is an important
direction for future work.

 \section{Conclusions}
 This paper analyzed the joint impact of low-resolution ADCs and limited 
fronthaul quantization in the uplink of a wideband cell-free massive 
MIMO--OFDM system. Unlike conventional approaches that suffer from 
double quantization---first in the time domain due to ADCs and then in 
the frequency domain for fronthaul transmission—we proposed a 
quantization-aware strategy in which each AP reconstructs the clean 
time-domain signal prior to fronthaul compression. By formulating the 
problem using the structure of the OFDM modulation and the symbol 
constellation, we developed an efficient ADMM-based algorithm capable of 
estimating the unquantized signal while respecting the ADC quantization 
intervals.

Numerical results demonstrated that the proposed method significantly 
reduces the reconstruction error compared to a quantization-unaware 
benchmark, consistently across different fronthaul resolutions and 
constellation sizes. Although the study focused on per-AP signal 
reconstruction, the final data detection is performed at the CPU; 
designing advanced detection strategies based on the proposed fronthaul 
structure is an important direction for future work.

\bibliographystyle{IEEEtran}
\bibliography{IEEEabrv,refs}

\begin{thebibliography}{10}
\providecommand{\url}[1]{#1}
\csname url@samestyle\endcsname
\providecommand{\newblock}{\relax}
\providecommand{\bibinfo}[2]{#2}
\providecommand{\BIBentrySTDinterwordspacing}{\spaceskip=0pt\relax}
\providecommand{\BIBentryALTinterwordstretchfactor}{4}
\providecommand{\BIBentryALTinterwordspacing}{\spaceskip=\fontdimen2\font plus
\BIBentryALTinterwordstretchfactor\fontdimen3\font minus
  \fontdimen4\font\relax}
\providecommand{\BIBforeignlanguage}[2]{{%
\expandafter\ifx\csname l@#1\endcsname\relax
\typeout{** WARNING: IEEEtran.bst: No hyphenation pattern has been}%
\typeout{** loaded for the language `#1'. Using the pattern for}%
\typeout{** the default language instead.}%
\else
\language=\csname l@#1\endcsname
\fi
#2}}
\providecommand{\BIBdecl}{\relax}
\BIBdecl

\bibitem{Ngo2017b}
H.~Q. Ngo, A.~Ashikhmin, H.~Yang, E.~G. Larsson, and T.~L. Marzetta,
  ``Cell-free {Massive} {MIMO} versus small cells,'' \emph{{IEEE} Trans.
  Wireless Commun.}, vol.~16, no.~3, pp. 1834--1850, 2017.

\bibitem{cell-free-book}
\BIBentryALTinterwordspacing
\"{O}zlem Tugfe~Demir, E.~Bj\"{o}rnson, and L.~Sanguinetti, ``Foundations of
  user-centric cell-free massive mimo,'' \emph{Foundations and Trends® in
  Signal Processing}, vol.~14, no. 3-4, pp. 162--472, 2021. [Online].
  Available: \url{http://dx.doi.org/10.1561/2000000109}
\BIBentrySTDinterwordspacing

\bibitem{ngo2024ultradense}
H.~Q. Ngo, G.~Interdonato, E.~G. Larsson, G.~Caire, and J.~G. Andrews,
  ``Ultradense cell-free massive {MIMO} for {6G}: Technical overview and open
  questions,'' \emph{Proceedings of the IEEE}, 2024.

\bibitem{8891922}
H.~Masoumi and M.~J. Emadi, ``Performance analysis of cell-free massive {MIMO}
  system with limited fronthaul capacity and hardware impairments,'' \emph{IEEE
  Transactions on Wireless Communications}, vol.~19, no.~2, pp. 1038--1053,
  2020.

\bibitem{8756265}
X.~Hu, C.~Zhong, X.~Chen, W.~Xu, H.~Lin, and Z.~Zhang, ``Cell-free massive
  {MIMO} systems with low resolution {ADCs},'' \emph{IEEE Transactions on
  Communications}, vol.~67, no.~10, pp. 6844--6857, 2019.

\bibitem{8678745}
G.~Femenias and F.~Riera-Palou, ``Cell-free millimeter-wave massive {MIMO}
  systems with limited fronthaul capacity,'' \emph{IEEE Access}, vol.~7, pp.
  44\,596--44\,612, 2019.

\bibitem{9123382}
------, ``Fronthaul-constrained cell-free massive {MIMO} with low resolution
  {ADC}s,'' \emph{IEEE Access}, vol.~8, pp. 116\,195--116\,215, 2020.

\bibitem{demirWiopt}
{\"{O}}.~T. Demir, A.~M. Elbir, and E.~Björnson, ``Cell-free massive
  {MIMO-OFDM} with low-resolution {ADCs},'' in \emph{2025 23rd International
  Symposium on Modeling and Optimization in Mobile, Ad Hoc, and Wireless
  Networks (WiOpt)}, 2025, pp. 1--7.

\bibitem{3GPP5G}
\emph{{5G}; Study on channel model for frequencies from 0.5 to 100 GHz (Release
  14)}.\hskip 1em plus 0.5em minus 0.4em\relax {3GPP} {TR} 38.901, Jan. 2018.

\bibitem{Saleh1987}
A.~A. Saleh and R.~A. Valenzuela, ``A statistical model for indoor multipath
  propagation,'' \emph{{IEEE} J. Sel. Areas Commun.}, vol.~5, no.~2, pp.
  128--137, 1987.

\bibitem{bjornson2024introduction}
E.~Bj{\"o}rnson and {\"O}.~T. Demir, \emph{Introduction to multiple antenna
  communications and reconfigurable surfaces}.\hskip 1em plus 0.5em minus
  0.4em\relax Now Publishers, Inc., 2024.

\end{thebibliography}

\end{document}